\documentclass[twocolumn,prd,nofootinbib]{revtex4}
\usepackage{amsmath}
\usepackage{amssymb}

\begin{document}

\title{Friedmann branes with variable tension}
\author{L\'{a}szl\'{o} \'{A}rp\'{a}d Gergely}
\affiliation{Department of Theoretical Physics, University of Szeged, Tisza Lajos krt
84-86, Szeged 6720, Hungary}
\affiliation{Department of Experimental Physics, University of Szeged, D\'{o}m T\'{e}r 9,
Szeged 6720, Hungary}
\email{gergely@physx.u-szeged.hu}
\affiliation{Department of Applied Science, London South Bank University, 103 Borough
Road, London SE1 0AA, UK}
\email{gergelyl@lsbu.ac.uk}
\date{\today}

\begin{abstract}
We introduce brane-worlds with non-constant tension, strenghtening the
analogy with fluid membranes, which exhibit a temperature-dependence
according to the empirical law established by E\"{o}tv\"{o}s. This new
degree of freedom allows for evolving gravitational and cosmological
constants, the latter being a natural candidate for dark energy. We
establish the covariant dynamics on a brane with variable tension in full
generality, by considering asymmetrically embedded branes and allowing for
non-standard model fields in the 5-dimensional space-time. Then we apply the
formalism for a perfect fluid on a Friedmann brane, which is embedded in a
5-dimensional charged Vaidya-Anti de Sitter space-time.
\end{abstract}

\startpage{1}
\endpage{}
\maketitle

\section{Introduction}

The hierarchy problem and the failure to quantize gravity in the way
other interactions were quantized are but two of the symptoms indicating the need
to break out from the established framework of field theories and general
relativity towards a more fundamental theory. Current attempts, like string
/ M-theory require additional spatial dimensions in addition to the 3+1
dimensional space-time. These additional dimensions are compact, however one
of them can be extended (but warped) in the scenario introduced by Randall
and Sundrum \cite{RS2}. Generalizations of this early model, allowing for a
curved 3+1 space-time with matter (the brane), embedded in curved
5-dimensional (5D) background \cite{SMS} have been developed (for a review
see \cite{MaartensLivRev}). By lifting the symmetry of the embedding and for
generic sources in 5D, the dynamics was worked out in detail in \cite{Decomp}%
. This work generalized previous discussions with asymmetric embedding \cite%
{asymmetry}.

For such co-dimension one brane-worlds the gravitational variables (namely
the induced metric and the extrinsic curvature) should satisfy the Israel
junction conditions \cite{Israel}. Standard model matter fields can be
introduced on the brane by the Lanczos equation \cite{Lanczos}, which
establishes a connection between the jump of the extrinsic curvature and the
brane energy-momentum tensor; similarly how various components of
the electromagnetic field exhibit a jump across surfaces with distributional
charge or current densities. This mechanism works only for co-dimension one,
therefore the generalization of these brane-worlds with arbitrary Riemannian
curvature to higher co-dimension seems far from straightforward.

In the context of co-dimension one brane-worlds black holes were found (in
the static \cite{tidalRN} and rotating \cite{Aliev} cases, or in
cosmological context \cite{BraneSwissCheese}). Gravitational collapse \cite%
{BGM}-\cite{BraneOppSnyder} together with various stellar models \cite{GM}-%
\cite{Ovalle} were also studied. The possibility that brane-world effects
can replace dark matter in galactic dynamics \cite{HarkoRC} and the dynamics
of clusters of galaxies \cite{HarkoClusters} were also considered. The
deflection of light was computed to second order accuracy \cite{GeDa} and a
confrontation with Solar System tests has been done \cite{BohmerHarkoLobo}.

Equally interesting applications of brane-worlds arise in cosmology.
Modifications to the evolution of the early universe were discussed in \cite%
{BDEL}. From the thermal radiation of an initial very hot brane even a black
hole can condensate in the extra dimension \cite{ChKN}. This black hole, in
turn modifies the Weyl curvature and back-reacts onto the intrinsic
curvature (and consequently the gravitation) of the brane. Structure
formation has been considered to some extent in \cite{PalStructure}. However
the equations do not close on the brane, therefore (despite progress made 
\cite{perturb}) a full perturbation formalism on the brane is not yet
available. Such a formalism would be necessary in order to discuss density
perturbations in relation with the Cosmic Microwave Background (CMB) and
structure formation in full generality. Nucleosynthesis constraints \cite%
{Nucleosynthesis} and confrontation with distant Ia type supernova data \cite%
{supernova} have been employed in order to establish the range of various
brane-world model parameters.

A key ingredients in brane-world theories is a positive brane tension. Its
value should be enormously high, such that when fine-tuned to the 5D
(negative) cosmological constant, they add up to a small positive
cosmological constant in 4 dimensions (4D). There are various lower limits
established for the brane tension from measurements on the gravitational
constant \cite{tabletop}, from the requirement that brane-world effects are
already weak at nucleosynthesis \cite{nucleosynthesis} and from
astrophysical considerations \cite{GM} (for a review of these topics see 
\cite{GK}).

The brane tension is the analogue of the tension of a fluid membrane, which
however is not a constant. In 1886 E\"{o}tv\"{o}s established an empirical
law for the temperature dependence of fluid membrane tension $\lambda
_{fluid}$, know today as E\"{o}tv\"{o}s' law \cite{Eotvos}. According to
this 
\begin{equation}
\lambda _{fluid}=K\left( T_{c}-T\right) ~,  \label{EotvosLaw}
\end{equation}%
$K$ being a constant and $T_{c}\,$\ a critical temperature representing the
highest temperature for which the membrane exists.

During cosmological evolution, the temperature of the brane (given by the
CMB) changes drastically from very high values to nowadays $2.7$ K. The
question naturally arises, why should the brane tension stay constant during
this spectacular evolution?

In this paper we lift the assumption of constancy of the brane tension. We
derive the co-dimension one brane-world dynamics with variable brane
tension. In Section II we decompose the 5D Einstein equations with respect
to the brane and we perform those transformations which lead to the
effective Einstein equation. We give there the complete dynamics in the most
generic case of an asymmetric embedding and arbitrary 5D sources. This is
given by the effective Einstein, Codazzi and twice contracted Gauss
equations on the brane. The most interesting applications of the developed
formalism would be for cosmological and black hole branes.

We consider cosmological branes in Section III, where we specialize our
results for a Friedmann brane with perfect fluid. We derive the generalized
Raychaudhuri and Friedmann equations, and also give the energy balance
equation, the twice-contracted Bianchi identity and the Lanczos equation for
this case.

It has been proved that the 5D vacuum space-time should be either
Schwarzschild - Anti de Sitter \cite{BCG}, \cite{BraneBlackHole} or its
horizon metric \cite{EinBrane}, \cite{EinBrane3}. In Section IV however we
consider the most general 5D space-time with radiation and electromagnetic
field which admits Friedmann branes in any point, the charged Vaidya-Anti de
Sitter (VAdS5) space-time. Such models with 5D radiation were considered
before in \cite{ChKN}, \cite{GK} and \cite{LSR}-\cite{HarkoInflation}.) We
analyze the geometry, the sources, the embedding and the dynamics,
represented by the Friedmann, Raychaudhuri and energy balance equations.
Finally we discuss the implications of the model and we summarize our
findings in Section V.

The formalism developed in this paper generalizes the results of Ref. \cite%
{Decomp} for variable brane tension. We also find that the formalism
developed in \cite{Decomp} applies only when pieces of interior charged
VAdS5 space-time regions are glued together along the brane, whereas the
results of the present paper stand for the more generic case when either
interior, or exterior regions are present on both sides of the brane.

Throughout the paper a tilde distinguishes the quantities defined on the
5-dimensional space-time, the only exception under this notation being the normal $n$ to the leaves
of the foliation. Its norm is $n^{c}n_{c}=1$. Latin indices represent
abstract indices running from $0$ to $5$. Vector fields in Lie-derivatives
are represented by boldface characters. An overbar denotes the average taken
over the left (L) and right (R) parts of the brane (with the exception of $%
\overline{L}_{ab}$, which is rather constructed from averaged quantities); $%
\Delta $ denotes the difference taken between the right and the left values
of a quantity and the superscript $^{TF}$ denotes trace-free (the trace
being formed with the brane metric).

\section{Brane-covariant gravitational dynamics}

Following \cite{Decomp} we summarize the most generic form of the equations
characterizing the evolution of gravitation in a co-dimension one
brane-world, allowing for asymmetric embedding and non-standard model
sources in 5D. For this we write the 5D metric $\widetilde{g}_{ab}$ in terms
of the brane normal $n^{a}=\left( \partial /\partial y\right) ^{a}$ and the
induced metric $g_{ab}$ as: 
\begin{equation}
\widetilde{g}_{ab}=n_{a}n_{b}+g_{ab}\ .  \label{tildeg0}
\end{equation}%
The 5D geometry evolves according to the 5D Einstein equation%
\begin{equation}
\widetilde{G}_{ab}=\widetilde{\kappa }^{2}\left[ -\widetilde{\Lambda }%
\widetilde{g}_{ab}+\widetilde{T}_{ab}+\tau _{ab}\delta \left( y-y_{b}\right) %
\right] \text{ .}  \label{5DEinstein}
\end{equation}%
where $\widetilde{\kappa }^{2}$ and $\widetilde{\kappa }^{2}\widetilde{%
\Lambda }$ are the 5D gravitational coupling constant and cosmological
constant, $\widetilde{T}_{ab}$ the regular part of the 5D energy-momentum
tensor representing the contribution of possible non-standard model fields
in 5D (like moduli fields or radiation from off-brane sources) and $\tau
_{ab}$ is the distributional part localized on the brane (at $y=y_{b}$),
obeying $\tau _{ik}n^{i}=0$. Usually the brane energy-momentum tensor $\tau
_{ab}$ is further decomposed as 
\begin{equation}
\tau _{ab}=-\lambda g_{ab}+T_{ab}~,  \label{tau}
\end{equation}%
where $\lambda $ is the brane tension and $T_{ab}$ represents ordinary
matter on the brane.

\subsection{The effective Einstein equation}

The 4D metric $g_{ab}$ evolves on the brane according to the effective
Einstein equation%
\begin{equation}
G_{ab}=-\Lambda g_{ab}+\kappa ^{2}T_{ab}+\widetilde{\kappa }^{4}S_{ab}-%
\overline{\mathcal{E}}_{ab}+\overline{L}_{ab}^{TF}+\overline{\mathcal{P}}%
_{ab}\ .  \label{modEgen}
\end{equation}%
As shown in \cite{Decomp}, this equation comes from a combination of the
scalar and trace-free part of the tensorial projections of the 5D Einstein
equation (\ref{5DEinstein}); the definition of the electric part $\mathcal{E}%
_{ac}=\widetilde{C}_{abcd}n^{b}n^{d}$ of the 5D Weyl curvature; finally from
averaging over the two sides of the brane, which requires to apply the
junction conditions.

The 4D gravitational coupling 'constant' $\kappa ^{2}$ is given by 
\begin{equation}
6\kappa ^{2}=\widetilde{\kappa }^{4}\lambda \ .  \label{kappa2def}
\end{equation}%
For attracting gravity the brane tension should be positive $\lambda >0$.

The source term $S_{ab}$ denotes a quadratic expression in the brane
energy-momentum tensor $T_{ab}$: 
\begin{equation}
S_{ab}=\frac{1}{4}\Biggl[-T_{ac}^{\ }T_{b}^{c}+\frac{T}{3}T_{ab}-\frac{g_{ab}%
}{2}\left( -T_{cd}^{\ }T^{cd}+\frac{T^{2}}{3}\right) \Biggr]\ .  \label{S}
\end{equation}%
(The first four source terms of Eq. (\ref{modEgen}) were first derived for a
symmetric embedding in \cite{SMS}.)

The possible asymmetric embedding is characterized by the tensor 
\begin{equation}
\overline{L}_{ab}=\overline{K}_{ab}\overline{K}-\overline{K}_{ac}\overline{K}%
_{b}^{c}-\frac{g_{ab}}{2}\left( \overline{K}^{2}-\overline{K}_{cd}\overline{K%
}^{cd}\right) \ ,  \label{L}
\end{equation}%
(for a symmetric embedding $\overline{K}_{ab}=0$, thus $\overline{L}_{ab}=0$%
) with trace%
\begin{equation}
\overline{L}=\overline{K}_{ab}\overline{K}^{ab}-\overline{K}^{2}~,
\end{equation}
and trace-free part 
\begin{equation}
\overline{L}_{ab}^{TF}=\overline{K}_{ab}\overline{K}-\overline{K}_{ac}%
\overline{K}_{b}^{c}+\frac{\overline{L}}{4}g_{ab}~.
\end{equation}

Finally $\overline{\mathcal{P}}_{ab}$ is given by the pull-back to the brane
of the energy-momentum tensor characterizing possible non-standard model
fields (scalar, dilaton, moduli, radiation of quantum origin) living in 5D: 
\begin{equation}
\overline{\mathcal{P}}_{ab}=\frac{2\widetilde{\kappa }^{2}}{3}\overline{%
\left( g_{a}^{c}{}g_{b}^{d}{}\widetilde{T}_{cd}\right) ^{TF}}\ ,  \label{P}
\end{equation}%
which is traceless by definition.

Another projection $n^{c}n^{d}\widetilde{T}_{cd}$ of these 5D sources
appears in the brane cosmological 'constant' $\Lambda $, which in general
can vary both due to non-standard model fields and due to the asymmetric
embedding through $\overline{L}$%
\begin{equation}
\Lambda =\Lambda _{0}-\frac{\overline{L}}{4}-\frac{\widetilde{\kappa }^{2}}{2%
}\overline{\left( n^{c}n^{d}\widetilde{T}_{cd}\right) }\ ,  \label{Lambdadef}
\end{equation}%
with $\Lambda _{0}$ given as 
\begin{equation}
2\Lambda _{0}=\kappa ^{2}\lambda +\widetilde{\kappa }^{2}\overline{%
\widetilde{\Lambda }}~.  \label{Lambda0}
\end{equation}%
For varying $\lambda $ even $\Lambda _{0}$ fails to be a constant.

The effective Einstein equation does not represent a closed system. Indeed,
among its sources we find embedding variables $\overline{L}_{ab}^{TF}$ and $%
\overline{L}$, the non-local term $\mathcal{E}_{ac}$ as well as the $%
n^{c}n^{d}\widetilde{T}_{cd}$ and $g_{a}^{c}{}g_{b}^{d}{}\widetilde{T}_{cd}$
projections of 5D sources (their evolution being inter-twined with the
evolution of the 5D metric, again requires non-local knowledge of
gravitational dynamics).

\subsection{Difference equations}

We can also form the difference over the two sides of the brane for both the
scalar and the trace-free part of the tensorial projections of the 5D
Einstein equation (\ref{5DEinstein}), obtaining: 
\begin{subequations}
\label{syst1}
\begin{gather}
-\lambda \overline{K}+T_{ab}\overline{K}^{ab}=\Delta \left( n^{a}n^{b}%
\widetilde{T}_{ab}\right) -\Delta \widetilde{\Lambda }\ ,  \label{trm1a} \\
\Delta \mathcal{E}_{ab}=\frac{2\widetilde{\kappa }^{2}}{3}\Delta \left(
g_{a}^{c}{}g_{b}^{d}{}\widetilde{T}_{cd}\right) ^{TF}  \notag \\
-\widetilde{\kappa }^{2}\left[ \overline{K}T_{ab}+\frac{T}{3}\overline{K}%
_{ab}+\frac{2}{3}\lambda \overline{K}_{ab}-2\overline{K}_{(a}^{c}T_{b)c}%
\right] ^{TF}\ .  \label{trlessm1}
\end{gather}%
For given brane and 5D sources, the first of these equations represents a
constraint on the embedding. The second equation gives $\Delta \mathcal{E}%
_{ab}$ in terms of embedding, brane and 5D sources.

\subsection{The Codazzi equation}

The vectorial projection of the of the 5D Einstein equation (\ref{5DEinstein}%
) gives a constraint on the brane variables $\left( g_{ab},~K_{ab}\right) $,
known as the Codazzi equation. When averaging and subtracting the
corresponding equations taken on the two sides of the brane we obtain 
\end{subequations}
\begin{subequations}
\label{system}
\begin{eqnarray}
\nabla _{c}\overline{K}_{a}^{c}-\nabla _{a}\overline{K} &=&\widetilde{\kappa 
}^{2}\overline{\left( g_{a}^{c}{}n^{d}\widetilde{T}_{cd}\right) }\ ,
\label{vpCodazzi} \\
\nabla _{c}\tau _{a}^{c} &=&-\Delta \left( g_{a}^{c}{}n^{d}{}\widetilde{T}%
_{cd}\right) \ .  \label{vmCodazzi}
\end{eqnarray}%
The first of these, the averaged Codazzi equation is a constraint on the
embedding and 5D sources (for symmetric embedding only for the latter),
while the second, the difference Codazzi equation gives the energy balance
on the brane. Written in more detail, we obtain: 
\end{subequations}
\begin{equation}
\nabla _{c}T_{a}^{c}=\nabla _{a}\lambda -\Delta \left( g_{a}^{c}{}n^{d}{}%
\widetilde{T}_{cd}\right) \ .  \label{en_balance}
\end{equation}%
This is the first equation, which is modified with respect to the constant $%
\lambda $ case by our assumption of a non-constant brane tension. It tells
that the energy balance on the brane can be changed in two ways: (i) due to
specific non-standard model fields in 5D, like radiation (this has been
explored in \cite{ChKN}-\cite{Langlois}) and (ii) due to the varying brane
tension.

\subsection{The twice-contracted Gauss equation}

As shown in \cite{Decomp}, the scalar projection of the 5D Einstein equation
(\ref{5DEinstein}) is by construction the trace of the effective Einstein
equation (\ref{modEgen}). Then the trace of the tensorial projection gives
the remaining independent scalar equation, equivalent to the
twice-contracted Gauss equation or (after eliminating $R$ with the help of
the trace of Eq. (\ref{modEgen})) to%
\begin{equation}
E=\widetilde{\kappa }^{2}\left( n^{a}n^{b}\widetilde{T}_{ab}-\frac{%
\widetilde{T}}{3}+\frac{2\widetilde{\Lambda }}{3}\right) \ .  \label{Etrace}
\end{equation}%
We have denoted by $E$ the trace of 
\begin{equation}
E_{ab}=K_{ac}K_{b}^{c}-\mathcal{L}_{\mathbf{n}}K_{ab}+\nabla _{b}\alpha
_{a}-\alpha _{b}\alpha _{a}\ ,  \label{Eab}
\end{equation}%
carrying information about the off-brane evolution of $K_{ab}$. Here $\alpha
^{b}=n^{c}\widetilde{\nabla }_{c}n^{b}=g_{d}^{b}\alpha ^{d}$ is the
curvature of the congruence $n^{a}$.

\subsection{The Lanczos equation}

The Lanczos equation was already employed in the averaged equations above,
but is also useful for monitoring the off-brane gravitational sector.

The junction conditions across the brane, written in a covariant way by
Israel \cite{Israel}, include (a) the continuity of the induced metric $%
g_{ab}=g_{ab}^{R}=g_{ab}^{L}$, and (b) the Lanczos equation \cite{Lanczos},
which establishes the connection between the jump in the extrinsic curvature
and the energy-momentum tensor on the brane: 
\begin{equation}
\Delta K_{ab}=-\widetilde{\kappa }^{2}\left( \tau _{ab}-\frac{\tau }{3}%
g_{ab}\right) \ ,  \label{Lanczos}
\end{equation}%
or equivalently 
\begin{equation}
-\widetilde{\kappa }^{2}\tau _{ab}=\Delta K_{ab}-g_{ab}\Delta K\ .
\label{Lanczos1}
\end{equation}%
The extrinsic curvature on the two sides is given by 
\begin{equation}
2K_{ab}^{R,L}=2\overline{K}_{ab}\pm \Delta K_{ab}\ .  \label{Kab}
\end{equation}%
Here the second term is determined by the brane energy-momentum tensor
through the Lanczos equation, while the first should be a solution of the
first Codazzi equation (\ref{vpCodazzi}).

\subsection{Off-brane gravitational evolution}

The brane gravitational variables are the induced metric $g_{ab}$ and the
extrinsic curvature $K_{ab}$. They can be evolved along the off-brane normal 
$n^{a}$ in the following way. Eq. (\ref{Kab}) gives the off-brane evolution
of $g_{ab}$ cf. the definition of the extrinsic curvature%
\begin{equation}
\mathcal{L}_{\mathbf{n}}g_{ab}=2K_{ab}\ ,  \label{Lieg}
\end{equation}%
$\mathcal{L}_{\mathbf{n}}$ denoting the brane-projected Lie-derivative along
the brane normal.

The off-brane evolution of $K_{ab}$ can be found by rewriting Eq. (\ref{Eab}%
) in the form%
\begin{equation}
\mathcal{L}_{\mathbf{n}}K_{ab}=-\left( E_{ab}^{TF}+\frac{E}{4}g_{ab}\right)
+K_{ac}K_{b}^{c}+\nabla _{b}\alpha _{a}-\alpha _{b}\alpha _{a}\ ,
\label{LieK}
\end{equation}%
where $E$ can be expressed in terms of the 5D sources cf. Eq. (\ref{Etrace}%
). The trace-free part of $E_{ab}$ is found from the definition of $\mathcal{%
E}_{ab}$ and the trace-free part of the tensorial projection of the 5D
Einstein equation \cite{Decomp} as 
\begin{equation}
E_{ab}^{TF}=\mathcal{E}_{ab}-\frac{\widetilde{\kappa }^{2}}{3}\left[
g_{a}^{c}{}g_{b}^{d}\widetilde{T}_{cd}+\tau _{ab}\delta \left( y\right) %
\right] ^{TF}\ .  \label{eps1}
\end{equation}%
Here $\mathcal{E}_{ab}$ can be further decomposed as $\mathcal{E}_{ab}^{R,L}=%
\overline{\mathcal{E}}_{ab}\pm \Delta \mathcal{E}_{ab}/2$, with $\Delta 
\mathcal{E}_{ab}$ given by the difference equation (\ref{trlessm1}) and $%
\overline{\mathcal{E}}_{ab}$ by the trace-free part of the effective Einstein equation
(\ref{modEgen}).

\subsection{The brane Bianchi identity}

The covariant divergence of the effective Einstein equation gives the
twice-contracted Bianchi identity in $4$ dimensions, from which the
longitudinal part of $\left( \overline{\mathcal{E}}_{ab}-\overline{L}%
_{ab}^{TF}-\overline{\mathcal{P}}_{ab}\right) $ can be expressed as:%
\begin{gather}
\nabla ^{a}\left( \overline{\mathcal{E}}_{ab}-\overline{L}_{ab}^{TF}-%
\overline{\mathcal{P}}_{ab}\right) =  \notag \\
\frac{\nabla _{b}\overline{L}}{4}+\frac{\widetilde{\kappa }^{2}}{2}\nabla
_{b}\overline{\left( n^{c}n^{d}\widetilde{T}_{cd}\right) }-\kappa ^{2}\Delta
\left( g_{b}^{c}{}n^{d}{}\widetilde{T}_{cd}\right)  \notag \\
+\frac{\widetilde{\kappa }^{4}}{4}\left( T_{b}^{a}-\frac{T}{3}%
g_{b}^{a}\right) \Delta \left( g_{a}^{c}{}n^{d}{}\widetilde{T}_{cd}\right) 
\notag \\
+\frac{\widetilde{\kappa }^{4}}{4}\left[ 2T^{ac}\nabla _{\lbrack
b}T_{a]c}^{\ }+\frac{1}{3}\left( T_{ab}\nabla ^{a}T-T\nabla _{b}T\right) %
\right]  \notag \\
-\frac{\widetilde{\kappa }^{4}}{12}\left( T_{b}^{a}-Tg_{b}^{a}\right) \nabla
_{a}\lambda \ .  \label{Bianchi}
\end{gather}%
In deriving this identity we have used the relations $\nabla ^{a}\kappa
^{2}=\left( \kappa ^{2}/\lambda \right) \nabla ^{a}\lambda $ and $\nabla
^{a}\Lambda _{0}=\kappa ^{2}\nabla ^{a}\lambda $, deductible from the
definitions of $\kappa ^{2}$ and $\Lambda _{0}$, Eqs. (\ref{kappa2def}) and (%
\ref{Lambda0}), respectively; together with the energy balance equation (\ref%
{en_balance}). We note that there are manifest contributions due to the
varying brane tension in the covariant divergence of the effective Einstein
equation, such that (a) the term $\nabla ^{a}\left( \kappa ^{2}\right)
T_{ab} $, (b) the term $-g_{ab}\nabla ^{a}\Lambda $ $=...+\kappa ^{2}\nabla
_{b}\lambda $, however other $\nabla ^{a}\lambda $ contributions arise when
we replace the covariant divergence of $T_{ab}$ through its expression (\ref%
{en_balance}). As a result the contribution (b) is canceled, and the
coefficient of the contribution (a) is changed. The whole contribution to
the varying brane tension is encompassed in the last term of Eq. (\ref%
{Bianchi}). The cosmological implications of the twice-contracted Bianchi
identity will be exploited in the next section.

We note that terms due to the variable tension appear only in Eqs. (\ref%
{en_balance}) and (\ref{Bianchi}).

\section{Perfect fluid on Friedmann brane}

In this section we consider branes with cosmological symmetry (Friedmann
branes), containing perfect fluid, however we leave unspecified both the 5D
geometry and 5D sources, for possible future applications.

The metric on a Friedmann brane can be characterized covariantly as 
\begin{equation}
g_{ab}=-u_{a}u_{b}+a^{2}\left( \tau \right) h_{ab}\ ,  \label{f_metric}
\end{equation}%
where $a\left( \tau \right) $ is the scale factor, $\tau $ is cosmological
time and $h_{ab}$ is a 3-metric with \textit{constant} curvature (with
curvature index $k=1,0,-1$) of the maximally symmetric spacial slices of
constant $\tau $. The timelike congruence $u^{a}=\left( \partial /\partial
\tau \right) ^{a}$ obeys $u^{a}u_{a}=-1$ and $h_{ab}u^{a}=0$. From the first
condition $u_{b}\nabla _{a}u^{b}=0$ also follows. In a basis adapted to $%
u^{a}$ the vector $u^{b}\nabla _{b}u^{a}$ can be easily shown to vanish. As
the vanishing of any tensor is a basis-independent statement, it is
generally true that $u^{b}\nabla _{b}u^{a}=0$.

We define the time-derivative with respect to $\tau $ (and denote it by a
dot) as the Lie-derivative in the $u^{a}$ direction, projected into the
hypersurface perpendicular to $u^{a}$ (of constant $\tau $). The condition $%
\dot{h}_{ab}=0$ then gives 
\begin{equation}
u^{c}\nabla _{c}h_{ab}=-\frac{1}{a^{2}}\left( \nabla _{a}u_{b}+\nabla
_{b}u_{a}\right) \ ,
\end{equation}%
with the trace 
\begin{equation}
\nabla _{a}u^{a}=3\frac{\dot{a}}{a}~.  \label{divu}
\end{equation}

The brane tension should vary only in a way to obey cosmological symmetries,
thus it can depend only on the time $\tau $, such that $\lambda =\lambda
\left( \tau \right) $. The brane perfect fluid is characterized by the
energy-momentum tensor 
\begin{equation}
T_{ab}=\rho \left( \tau \right) u_{a}u_{b}+p\left( \tau \right) a^{2}h_{ab}\
,  \label{f_enmom}
\end{equation}%
with $u^{a}$ its $4$-velocity. Spatial isotropy and homogeneity implies $%
h_{ab}\nabla ^{b}a=h_{ab}\nabla ^{b}\rho =h_{ab}\nabla ^{b}p=h_{ab}\nabla
^{b}\lambda =0$, thus for any of the functions $f=\left( a,~\rho
,~p,~\lambda \right) $ we have 
\begin{equation}
\nabla _{a}f=g_{a}^{b}\nabla _{b}f=-u_{a}\dot{f}~.  \label{fdot}
\end{equation}%
A similar relation applies for any other function of $\tau $.

The quadratic term (\ref{S}) for the perfect fluid takes the form: 
\begin{equation}
\widetilde{\kappa }^{4}S_{ab}=\kappa ^{2}\frac{\rho }{\lambda }\left[ \frac{%
\rho }{2}u_{a}u_{b}+\left( \frac{\rho }{2}+p\right) a^{2}h_{ab}\right] \ .
\label{f_S}
\end{equation}%
For the rest of the tracefree source terms we introduce the effective energy
density $U$ as: 
\begin{equation}
-\overline{\mathcal{E}}_{ab}+\overline{L}_{ab}^{TF}+\overline{\mathcal{P}}%
_{ab}=\kappa ^{2}U\left( u_{a}u_{b}+\frac{a^{2}}{3}h_{ab}\right) \ .
\label{U}
\end{equation}%
$U$ is composed of non-local (Weyl), embedding and 5D matter contributions.

The Einstein tensor of the metric (\ref{f_metric}) is 
\begin{equation}
G_{ab}=3\frac{\dot{a}^{2}+k}{a^{2}}u_{a}u_{b}-\left[ 2a\ddot{a}+\dot{a}^{2}+k%
\right] h_{ab}\ .  \label{f_ein}
\end{equation}%
Then the non-trivial projections of the effective Einstein equation (\ref%
{modEgen}) combine to the generalized Friedmann and generalized Raychaudhuri
equations: 
\begin{gather}
3\frac{\dot{a}^{2}+k}{a^{2}}=\Lambda +\kappa ^{2}\left[ \rho \left( 1+\frac{%
\rho }{2\lambda }\right) +U\right] \ ,  \label{Fried} \\
6\frac{\ddot{a}}{a}=2\Lambda -\kappa ^{2}\left[ \rho \left( 1+\frac{2\rho }{%
\lambda }\right) +3p\left( 1+\frac{\rho }{\lambda }\right) +2U\right] \ .
\label{Raych}
\end{gather}

The energy balance equation (\ref{en_balance}) decouples into temporal and
spatial projections:%
\begin{eqnarray}
\dot{\rho}+3\frac{\dot{a}}{a}\left( \rho +p\right) &=&-\dot{\lambda}+\Delta
\left( u^{c}{}n^{d}{}\widetilde{T}_{cd}\right) ~,  \label{f_en_balance_t} \\
\Delta \left( h_{a}^{c}{}n^{d}{}\widetilde{T}_{cd}\right) &=&0\ .
\label{f_en_balance_sp}
\end{eqnarray}%
Note that the normal vectors on the two sides of the brane are $n_{R}=n$ and 
$n_{L}=-n$, therefore the second term on the right hand side of Eq. (\ref%
{f_en_balance_t}) can be non-vanishing even in the symmetric case. When both
terms on the right hand side of Eq. (\ref{f_en_balance_t}) vanish, the
physical fluid obeys a continuity equation. When the 5D sources obey $\Delta
\left( u^{c}{}n^{d}{}\widetilde{T}_{cd}\right) =0$, while the brane tension
is varying, Eq. (\ref{f_en_balance_t}) becomes a continuity equation for the
fluid with energy density $\rho +\lambda $ and pressure $p-\lambda $.

The twice-contracted Bianchi identity (\ref{Bianchi}) can be specified for the chosen cosmological setup by
employing Eqs. (\ref{kappa2def}), (\ref{f_metric}), (\ref{divu}), (\ref%
{f_enmom}), (\ref{U}), (\ref{f_en_balance_sp}) and the remark that in order
to fulfill the cosmological symmetries, the functions $U$, $\overline{L}$
and $\overline{\left( n^{c}n^{d}\widetilde{\Pi }_{cd}\right) }$ depend only
on $\tau $. We find that the space projection identically vanishes, while
the temporal projection gives%
\begin{gather}
\kappa ^{2}\left( \dot{U}+4U\frac{\dot{a}}{a}+U\frac{\dot{\lambda}}{\lambda }%
\right) =\left[ \frac{\overline{L}}{4}+\frac{\widetilde{\kappa }^{2}}{2}%
\overline{\left( n^{c}n^{d}\widetilde{T}_{cd}\right) }\right] ^{\mathbf{%
\skew{01} {\dot} {}}}  \notag \\
-\kappa ^{2}\left( 1+\frac{\rho }{\lambda }\right) \Delta \left(
u^{c}{}n^{d}{}\widetilde{T}_{cd}\right) \ .  \label{f_Bianchi}
\end{gather}%
Again, some manifest contributions from the twice-contracted Bianchi
identity (\ref{Bianchi}), like the one emerging from the last term $-\left( 
\widetilde{\kappa }^{4}/4\right) p\dot{\lambda}u_{b}$ have canceled with
contributions from the term quadratic in $T_{ab}$, when we employed the
energy-balance equation (\ref{f_en_balance_t}). As consequence the
twice-contracted Bianchi identity (\ref{f_Bianchi}) has only one explicit $%
\dot{\lambda}$-term, originating in the derivative of $\left( \kappa
^{2}U\right) $. The equation can be written in a somewhat simpler form by
introducing the new function $U_{0}\left( \tau \right) $ with an $a^{-4}\,$\
dependence factorized out 
\begin{equation}
U=U_{0}\left( \frac{a_{0}}{a}\right) ^{4}\ ,  \label{Usol1}
\end{equation}%
where $a_{0}$ is an integration constant, and by re-introducing the function 
$\Lambda $ cf. Eqs.(\ref{Lambdadef}) and (\ref{Lambda0}):

\begin{gather}
\kappa ^{2}\left( \frac{a_{0}}{a}\right) ^{4}\left( \dot{U}_{0}+U_{0}\frac{%
\dot{\lambda}}{\lambda }\right) +\dot{\Lambda}-\kappa ^{2}\dot{\lambda} 
\notag \\
=-\kappa ^{2}\left( 1+\frac{\rho }{\lambda }\right) \Delta \left(
u^{c}{}n^{d}{}\widetilde{T}_{cd}\right) \ .  \label{U0}
\end{gather}

We have verified that the twice-contracted Bianchi identity (\ref{U0}) can
also be deduced in a direct way by taking the time derivative of the
generalized Friedmann equation, then employing the Raychaudhuri equation (%
\ref{Raych}) and the energy-balance equation (\ref{f_en_balance_t}).
According to this remark it is obvious that the energy-balance equation (\ref%
{f_en_balance_t}) is \textit{not} a consequence of the Friedmann and
Raychaudhuri equations, as is in the standard cosmological model, unless Eq.
(\ref{U0}) is identically satisfied. This latter condition holds, provided
the functions $U_{0}$ and $\Lambda $ are compatible with the 5D Einstein
equation and the embedding of the brane is properly chosen. An example of
what this means is provided in the next Section.

Finally we write the Lanczos equation (\ref{Lanczos1}) for a perfect fluid
on cosmological brane as 
\begin{equation}
\Delta K_{ab}\!=\!-\frac{\widetilde{\kappa }^{2}}{3}\!\left[ \!\left( 2\rho
+3p-\lambda \right) \!u_{a}u_{b}+\left( \rho +\lambda \right) a^{2}h_{ab}%
\right] ~,  \label{f_Lanczos}
\end{equation}%
a relation useful in relating brane variables to the extrinsic curvature,
and consequently to the study of off-brane gravitational evolution.

In summary, cosmological evolution on the brane is given by a generalized
Friedmann equation (\ref{Fried}), a generalized Raychaudhuri equation (\ref%
{Raych}) and an energy-balance equation (\ref{f_en_balance_t}). After
employing Eq. (\ref{Usol1}), these equations contain two unspecified
functions $\Lambda $ and $U_{0}$ depending on non-local contributions, the
asymmetry of the embedding and non-standard model 5D fields.

\section{Charged 5D Vaidya-Anti de Sitter space-time}

In this section in addition to the assumption of a variable tension
Friedmann brane with cosmological fluid, employed in the previous section,
we also specify the 5D space-time and the 5D sources, together with the
possible asymmetry in the embedding. These enables us to fix the functions $%
\Lambda $ and $U$ (or $U_{0}$) in the generalized Friedmann and Raychaudhuri
equations. Fixing them in such a way that the twice-contracted Bianchi
relation is obeyed, the energy balance equation becomes a consequence.

We choose the most generic 5D space-time with electromagnetic field and
unpolarized radiation (treated in the geometrical optics approximation),
admitting the cosmological symmetries of the brane in each point. The
discussion generalizes the one of \cite{Decomp} for variable brane tension.

\subsection{The geometry}

The 5D space-time is the charged Vaidya solution (VAdS5). In
Eddington-Finkelstein type coordinates 
\begin{align}
d\widetilde{s}^{2}& =-f\left( v,r;k\right) dv^{2}+2\epsilon dvdr  \notag \\
& +r^{2}\left[ d\chi ^{2}+\mathcal{H}^{2}\left( \chi ;k\right) \left(
d\theta ^{2}+\sin ^{2}\theta d\phi ^{2}\right) \right] \ ,  \label{ChVAdS5}
\end{align}%
where $\epsilon =1$ applies for an outgoing null coordinate $v$ (with
ingoing $v=$constant lines), while $\epsilon =-1$ for ingoing $v$ (outgoing $%
v=$constant lines). The metric functions are 
\begin{equation}
\mathcal{H}\left( \chi ;k\right) =\left\{ 
\begin{array}{c}
\sin \chi \ ,\qquad k=1 \\ 
\chi \ ,\qquad k=0 \\ 
\sinh \chi \ ,\qquad k=-1%
\end{array}%
\right. \ 
\end{equation}%
($k$ being the curvature index of the constant curvature 3-metric $h_{ab}$),
and%
\begin{equation}
f\left( v,r;k\right) =k-\frac{1}{r^{2}}\left[ 2m\left( v\right) +\frac{%
\widetilde{\kappa }^{2}\widetilde{\Lambda }}{6}r^{4}-\frac{q^{2}\left(
v\right) }{r^{2}}\right] \ .  \label{ff}
\end{equation}%
The functions $m\left( v\right) $ and $q\left( v\right) $ are freely
specifiable. Depending on the value of the parameters, the metric can have
one or two horizons or no horizon at all (for a discussion of the simplest
such metric with $m=$constant and $q=0$ see \cite{EinBrane3}). \ Due to the
allowed asymmetry, two regions of possibly different VAdS5 space-times can
be glued together across the brane, such that\ the global 5D space-time can
contain a charged black hole on none, one or either side of the brane. We
classify the possible left ($L$) and right ($R$) regions with an index $\eta
_{I}$ (with $I=\overline{L,R}$) taking the value $1$ if the region contains$%
\ r=0$, and $0$ otherwise.

The VAdS5 metric is written covariantly as 
\begin{equation}
\widetilde{g}_{ab}=-u_{a}u_{b}+n_{a}n_{b}+r^{2}h_{ab}\ .  \label{ChVAdS5a}
\end{equation}%
Instead of the $\left( v,~r\right) $ coordinate chart sometimes it is more
convenient to use the coordinates $\left( \tau ,~y\right) $ adapted to the
4-velocity $u^{a}$ of the fluid and the (right-pointing) brane normal $%
n^{a}=\left( -1\right) ^{\sigma }\left( \partial /\partial y\right) ^{a}$.
The sign $\left( -1\right) ^{\sigma }$ was introduced in order to allow the
coordinate $y$ either to increase or decrease in the direction of the brane
normal, which we choose to be right-pointing. For an \textit{outgoing}
coordinate $y$\ this is assured by 
\begin{equation}
\sigma =\left\{ 
\begin{array}{c}
\eta _{R}\ ,\qquad \text{right region} \\ 
\eta _{L}+1\ ,\qquad \text{left region}%
\end{array}%
\right. \ .
\end{equation}%
We also note, that $u$ is time-like (thus it can be interpreted as
4-velocity) only above the horizons. The dual coordinate bases are related
as 
\begin{eqnarray}
dv &=&\dot{v}d\tau +v^{\prime }dy~,  \notag \\
dr &=&\dot{r}d\tau +r^{\prime }dy~,  \label{dual}
\end{eqnarray}%
where the dot and the prime represent derivatives with respect to $\tau $
and $y$, respectively$.$The vector bases then transform with the transposed
inverse matrix, thus we have 
\begin{eqnarray}
u^{.} &\equiv &\frac{\partial }{\partial \tau }=\dot{v}\frac{\partial }{%
\partial v}+\dot{r}\frac{\partial }{\partial r}~,  \notag \\
\left( -1\right) ^{\sigma }n^{.} &\equiv &\frac{\partial }{\partial y}%
=v^{\prime }\frac{\partial }{\partial v}+r^{\prime }\frac{\partial }{%
\partial r}~.  \label{nu}
\end{eqnarray}%
Outside horizons the unit negative norm of $u^{a}$ implies 
\begin{equation}
f\dot{v}=\epsilon \dot{r}+S_{1}\left( \dot{r}^{2}+f\right) ^{1/2}~,
\label{vdot}
\end{equation}%
with $S_{1}^{2}=1$. In contrast to Ref. \cite{Decomp}, we do not chose the positive
sign in front of the square root, but leave it unspecified for later
convenience. A simple computation shows 
\begin{equation}
\dot{v}^{-1}=-\epsilon \dot{r}+S_{1}\left( \dot{r}^{2}+f\right) ^{1/2}~,
\label{vdotm1}
\end{equation}%
therefore outside the horizon ($f>0$) the above two equations imply that the
sign of $\dot{v}$ is given by $S_{1}$, irrespective of the value of $%
\epsilon $.

The 1-form $u_{.}$ then becomes%
\begin{equation}
u_{.}\equiv g\left( u^{.},.\right) =-S_{1}\left( \dot{r}^{2}+f\right)
^{1/2}dv+\epsilon \dot{v}dr~.  \label{uform}
\end{equation}%
The condition $g\left( n^{.},u^{.}\right) =0$ gives%
\begin{equation}
r^{\prime }=\frac{\epsilon f\dot{v}-\dot{r}}{\dot{v}}v^{\prime }=\epsilon
S_{1}\left( \dot{r}^{2}+f\right) ^{1/2}\frac{v^{\prime }}{\dot{v}}~,
\label{rprime}
\end{equation}%
while from the normalization of $n^{a}$, by employing Eqs. (\ref{vdot}), (%
\ref{vdotm1}) and (\ref{rprime}) we obtain%
\begin{equation}
v^{\prime }=S_{2}\dot{v}~,  \label{vprime}
\end{equation}%
with $S_{2}^{2}=1$. Then the normal form becomes%
\begin{equation}
n_{.}\equiv g\left( n^{.},.\right) =\left( -1\right) ^{\sigma }\epsilon
S_{2}\left( -\dot{r}dv+\dot{v}dr\right) ~.  \label{nform}
\end{equation}%
Starting from Eqs. (\ref{dual}) and employing Eq. (\ref{vdot}) it is now
straightforward to show $u_{.}=-d\tau $ and $n_{.}=\left( -1\right) ^{\sigma
}dy$, these relations being independent of the choices of the signs $S_{1}$
and$\ S_{2}$.

The electric part with respect to the brane normal $n^{a}$ of the Weyl
tensor of the space-time (\ref{ChVAdS5}) can be found by a straightforward
calculation:%
\begin{eqnarray}
\mathcal{E}_{ab} &=&\frac{1}{4r^{2}}\left( r^{2}\frac{\partial ^{2}f}{%
\partial r^{2}}-2r\frac{\partial f}{\partial r}+2f-2k\right) \left(
u_{a}u_{b}+\frac{r^{2}}{3}h_{ab}\right)  \notag \\
&=&\frac{3\left[ 5q^{2}\left( v\right) -4m\left( v\right) r^{2}\right] }{%
2r^{6}}\left( u_{a}u_{b}+\frac{r^{2}}{3}h_{ab}\right) ~.
\end{eqnarray}%
From here the Weyl source term in the effective Einstein equation emerges as%
\begin{eqnarray}
-\overline{\mathcal{E}}_{ab} &=&\kappa ^{2}U^{Weyl}\left( u_{a}u_{b}+\frac{%
r^{2}}{3}h_{ab}\right) ~,  \notag \\
\kappa ^{2}U^{Weyl} &=&\frac{6\overline{m}}{r^{4}}-\frac{15}{2r^{6}}\left( 
\overline{q}^{2}+\frac{\left( \Delta q\right) ^{2}}{4}\right) ~.
\label{UWeyl}
\end{eqnarray}%
(We have used that for any quantity $h$ the average of its square is $%
\overline{h^{2}}=\overline{h}^{2}+\left( \Delta h\right) ^{2}/4$.)

\subsection{The 5D sources}

The source of the charged VAdS5 metric (\ref{ChVAdS5}) is the 5D
cosmological constant $\widetilde{\kappa }^{2}\widetilde{\Lambda }$,
together with a superposition of a radiation stream, which generates $%
m\left( v\right) $ and an electromagnetic field, responsible for $q\left(
v\right) $.

The electromagnetic field is characterized by the energy-momentum tensor: 
\begin{equation}
\widetilde{T}_{ab}^{EM}=\frac{3q^{2}\left( v\right) }{\widetilde{\kappa }%
^{2}r^{6}}\left( u_{a}u_{b}-n_{a}n_{b}+r^{2}h_{ab}\right) \ ,  \label{TEM}
\end{equation}%
and is generated by a null 5-potential $A_{a}=l_{a}q\left( v\right) /r^{2}$.

The radiation stream (null dust in the geometrical optics approximation) is
described by the energy-momentum tensor:%
\begin{equation}
\widetilde{T}_{ab}^{ND}=\frac{3\beta \left( v,r\right) }{\widetilde{\kappa }%
^{2}r^{3}}l_{a}l_{b}\ .  \label{TND}
\end{equation}%
Such radiation is ingoing (towards $r=0$) for $\epsilon =1$ and outgoing for 
$\epsilon =-1$.\textbf{\ }This means that the radiation flux can either
approach or leave the brane and it shouldn't necessarily be the same on the
two sides. From the definitions of $\epsilon _{I}$ and $\eta _{I}$ it can be
checked that the global sign $\epsilon _{I}\left( -1\right) ^{\eta _{I}}$ is
negative for any radiation leaving the brane and positive for any radiation
arriving to the brane.

In Eq. (\ref{TND}) $l$ is a null 1-form: 
\begin{equation}
l=dv=\dot{v}\left[ \left( -1\right) ^{\sigma }S_{2}n-u\right] \ ,
\end{equation}%
thus in terms of the fluid 4-velocity and brane normal the energy-momentum
tensor becomes 
\begin{equation}
\widetilde{T}_{ab}^{ND}=\frac{3\beta \dot{v}^{2}}{\widetilde{\kappa }%
^{2}r^{3}}\left[ n_{a}n_{b}+2\left( -1\right) ^{\sigma
+1}S_{2}u_{(a}n_{b)}+u_{a}u_{b}\right] \ .
\end{equation}%
The function $\beta \left( v,r\right) $ is related to the energy density of
radiation (with dimension of linear density of mass) and the 5D Einstein
equations imply%
\begin{equation}
\epsilon \beta =\frac{dm}{dv}-\frac{q}{r^{2}}\frac{dq}{dv}\ .  \label{beta}
\end{equation}

On the brane (where $r=a\left( \tau \right) $, see the following subsection)
the projection of the total energy-momentum tensor $\widetilde{T}_{ab}=%
\widetilde{T}_{ab}^{ND}+\widetilde{T}_{ab}^{EM}$ needed in the energy
balance equation (\ref{f_en_balance_t}) becomes%
\begin{equation}
u^{c}{}n^{d}{}\widetilde{T}_{cd}=\frac{3\epsilon \left( -1\right) ^{\sigma
}\beta \dot{v}^{2}}{\widetilde{\kappa }^{2}a^{3}}~,
\end{equation}%
such that%
\begin{eqnarray}
\dot{\rho}+3\frac{\dot{a}}{a}\left( \rho +p\right) &=&-\dot{\lambda}+\frac{3%
}{\widetilde{\kappa }^{2}a^{3}}\Delta \left[ \epsilon \left( -1\right)
^{\sigma }\beta \dot{v}^{2}\right]  \notag \\
&=&-\dot{\lambda}+\frac{3}{\widetilde{\kappa }^{2}a^{3}}\sum\limits_{I=L,R}%
\!\epsilon _{I}\left( -1\right) ^{\eta _{I}}\!\beta _{I}\dot{v}_{I}^{2}\ ,
\label{conti_V}
\end{eqnarray}%
The global sign $\epsilon _{I}\left( -1\right) ^{\eta _{I}}$ is negative if
radiation leaves the brane and positive if radiation is absorbed on the
brane.

Finally the source term $\overline{\mathcal{P}}_{ab}$ arising from $%
\widetilde{T}_{cd}$ becomes%
\begin{eqnarray}
\overline{\mathcal{P}}_{ab} &=&\kappa ^{2}U^{ch-rad}\left( u_{a}u_{b}+\frac{%
a^{2}}{3}h_{ab}\right) \ ,  \notag \\
\kappa ^{2}U^{ch-rad} &=&\frac{3}{2a^{3}}\overline{\beta \dot{v}^{2}}+\frac{%
3}{a^{6}}\left( \overline{q}^{2}+\frac{\left( \Delta q\right) ^{2}}{4}%
\right) ~.  \label{U5D}
\end{eqnarray}

\subsection{The embedding}

The brane is located at $y=$const, thus in the coordinates ($\tau ,~y$) its
movement is encoded only in the change of the coordinate $\tau $. Therefore
the embedding relations are $v=v\left( \tau \right) $ given by Eq. (\ref%
{vdot}) and $r=a\left( \tau \right) $. The latter simply allows to replace $%
r $ and $\dot{r}$\ with $a$ and $\dot{a}$ in all expressions of the previous
subsection, whenever they are specified on the brane. The normal $n$ and
tangent $u$ to the brane are given by Eqs. (\ref{nu}), and the induced
metric by Eq. (\ref{f_metric}). The extrinsic curvature can then be
calculated to give 
\begin{eqnarray}
K_{ab} &=&\left( -1\right) ^{\sigma +1}\epsilon S_{1}S_{2}\Biggl[\frac{2%
\ddot{a}+\frac{\partial f}{\partial a}-\epsilon \dot{v}^{2}\frac{\partial f}{%
\partial v}}{2\left( \dot{a}^{2}+f\right) ^{1/2}}u_{a}u_{b}  \notag \\
&&-\left( \dot{a}^{2}+f\right) ^{1/2}ah_{ab}\Biggr]\ .
\end{eqnarray}%
By introducing the notations%
\begin{equation}
2A_{I}=2\ddot{a}+\frac{\partial f_{I}}{\partial a}-\epsilon _{I}\dot{v}%
_{I}^{2}\frac{\partial f_{I}}{\partial v}\ ,  \label{AV}
\end{equation}%
\begin{equation}
B_{I}=\left( -1\right) ^{\eta _{I}}\left( \dot{a}^{2}+f_{I}\right) ^{1/2}\ ,
\label{B}
\end{equation}%
with $I=R$ (right region) or $L$ (left region), the jump and average of the
extrinsic curvature can be written as 
\begin{eqnarray}
\epsilon S_{1}S_{2}\Delta K_{ab}\!\!\! &=&\!\!\!-\!2\overline{\left( \frac{A%
}{B}\right) }u_{a}u_{b}+2\!\overline{B}~ah_{ab}\ ,  \label{delK} \\
2\epsilon S_{1}S_{2}\overline{K}_{ab}\!\!\! &=&\!\!\!-\Delta \!\left( \frac{A%
}{B}\right) ~u_{a}u_{b}+\Delta B~ah_{ab}~.  \label{aveK}
\end{eqnarray}%
From the expression (\ref{aveK}) of $\overline{K}_{ab}$ we obtain the
embedding contribution to $\Lambda $:%
\begin{equation}
\overline{L}=-\frac{3\Delta B}{2a}\left[ \Delta \!\left( \frac{A}{B}\right) +%
\frac{\Delta B}{a}\right] \ .  \label{Lbar}
\end{equation}%
and the embedding tracefree source term in the effective Einstein equation: 
\begin{eqnarray}
\overline{L}_{ab}^{TF} &=&\kappa ^{2}U^{emb}\left( u_{a}u_{b}+\frac{a^{2}}{3}%
h_{ab}\right)  \notag \\
\kappa ^{2}U^{emb}\! &=&\!\frac{3\Delta B}{8a}\left[ \frac{\Delta B}{a}%
\!-\!\Delta \!\left( \frac{A}{B}\right) \!\right] ~.  \label{Uembedding}
\end{eqnarray}%
Note that the sign ambiguity $\epsilon S_{1}S_{2}$ dropped out from the
above expressions of $\overline{L}_{ab}^{TF}$ and $\overline{L}$, as both
are quadratic in $\overline{K}_{ab}$.

\subsection{The generalized Friedmann and Raychaudhuri equations}

Comparing Eq. (\ref{delK}) with the Lanczos equation on the Friedmann brane,
Eq. (\ref{f_Lanczos}) we identify the averaged quantities $\!\overline{%
\left( A/B\right) }$ and $\!\overline{B}$ in terms of the brane tension and
fluid variables as%
\begin{eqnarray}
\epsilon S_{1}S_{2}\!\overline{\left( \frac{A}{B}\right) } &=&\!\frac{%
\widetilde{\kappa }^{2}}{6}\!\!\left( 2\rho +3p-\lambda \right) \!~,
\label{aveAoB} \\
\epsilon S_{1}S_{2}\overline{B} &=&-\!\frac{\widetilde{\kappa }^{2}}{6}%
\!\left( \rho +\lambda \right) a~.  \label{aveB}
\end{eqnarray}%
Next, by taking the square of Eq. (\ref{B})~and averaging, then by employing
Eqs. (\ref{kappa2def}) and (\ref{aveB}), we have

\begin{equation}
\frac{\kappa ^{2}}{6\lambda }\!\left( \rho +\lambda \right) ^{2}a^{2}+\frac{%
\left( \Delta B\right) ^{2}}{4}=\dot{a}^{2}+\overline{f}\ ,
\end{equation}%
Finally, by taking into account Eqs. (\ref{Lambda0}) and (\ref{ff}), we
obtain the Friedmann equation: 
\begin{eqnarray}
\frac{\dot{a}^{2}+k}{a^{2}} &=&\frac{\Lambda _{0}}{3}\!+\frac{\kappa
^{2}\rho }{3}\!\left( 1+\frac{\rho }{2\lambda }\right) +\frac{2\overline{m}}{%
a^{4}}-\frac{\overline{q}^{2}}{a^{6}}  \notag \\
&&+\frac{\left( \Delta B\right) ^{2}}{4a^{2}}-\frac{\left( \Delta q\right)
^{2}}{4a^{6}}~.  \label{Fr_V}
\end{eqnarray}%
Here $\Delta B$ can be calculated from the jump of the square of Eq. (\ref{B}%
), by employing the expressions of $f$ and $\overline{B}$, Eqs. (\ref{ff})
and (\ref{aveB}), and $\Delta \left( h^{2}\right) =2\overline{h}~\Delta h$
applied both for $B$ and $q$: 
\begin{equation}
\epsilon S_{1}S_{2}\Delta B=\frac{12a^{2}\Delta m-12\overline{q}\Delta q+%
\widetilde{\kappa }^{2}a^{6}\Delta \widetilde{\Lambda }}{2\!\widetilde{%
\kappa }^{2}a^{5}\left( \rho +\lambda \right) }\ .  \label{delB}
\end{equation}%
However we note that the combined sign $\epsilon S_{1}S_{2}$ does not appear
in the Friedmann equation, which contains $\left( \Delta B\right) ^{2}$.

From the definitions of the functions $A,~f$ and $\beta $, Eqs. (\ref{AV}), (%
\ref{ff}) and (\ref{beta}) we find the average and jump of $A$ as%
\begin{eqnarray}
\overline{A} &=&\ddot{a}+\frac{2\overline{m}}{a^{3}}-\frac{\widetilde{\kappa 
}^{2}\overline{\widetilde{\Lambda }}}{6}a-\frac{2}{a^{5}}\left( \!\overline{q%
}^{2}\!+\!\frac{\left( \Delta q\right) ^{2}}{4}\!\right) \!+\!\frac{%
\overline{\beta \dot{v}^{2}}}{a^{2}}\ ,  \label{aveA} \\
\Delta A &=&\frac{2\Delta m}{a^{3}}-\frac{\widetilde{\kappa }^{2}\Delta 
\widetilde{\Lambda }}{6}a-\frac{4\overline{q}\Delta q}{a^{5}}+\frac{\Delta
\left( \beta \dot{v}^{2}\right) }{a^{2}}\ .  \label{delA}
\end{eqnarray}%
The previously deduced expressions for $\Delta A$ and $\Delta B$ obey%
\begin{equation}
3\Delta A+\widetilde{\kappa }^{2}aC=\widetilde{\kappa }^{2}\left( \rho
+\lambda \right) \epsilon S_{1}S_{2}\Delta B\ ,
\end{equation}%
with%
\begin{equation}
C=\Delta \widetilde{\Lambda }+\frac{6\overline{q}\Delta q}{\widetilde{\kappa 
}^{2}a^{6}}-\frac{3\Delta \left( \beta \dot{v}^{2}\right) }{\widetilde{%
\kappa }^{2}a^{3}}\ .  \label{C}
\end{equation}%
The same relation can also be found from the definitions (\ref{AV}) and (\ref%
{B}). Next, from the definition of $\overline{\left( A/B\right) }$ we obtain
a second expression for $\overline{A}$:%
\begin{equation}
\overline{A}=\frac{1}{\overline{B}}\left[ \frac{\Delta A\Delta B}{4}+%
\overline{\left( \frac{A}{B}\right) }\left( \overline{B}^{2}-\frac{\left(
\Delta B\right) ^{2}}{4}\right) \right] ~.  \label{aveA1}
\end{equation}%
When writing \ up this relation in detail, by employing Eqs. (\ref{aveAoB}),
(\ref{aveB}), (\ref{delB}), (\ref{delA}), the global sign $\epsilon
S_{1}S_{2}$ drops out. Comparing the two expressions for $\overline{A}$, by
taking into account the relations (\ref{kappa2def}), (\ref{Lambda0}), the
Raychaudhuri equation emerges: 
\begin{eqnarray}
\frac{\ddot{a}}{a} &=&\frac{\Lambda _{0}}{3}-\!\frac{\kappa ^{2}}{6}\!\left[
\rho \left( 1+\frac{2\rho }{\lambda }\right) +3p\left( 1+\frac{\rho }{%
\lambda }\right) \right]  \notag \\
&&-\frac{2\overline{m}}{a^{4}}+\frac{2}{a^{6}}\left( \overline{q}^{2}+\frac{%
\left( \Delta q\right) ^{2}}{4}\right) -\frac{\overline{\beta \dot{v}^{2}}}{%
a^{3}}  \notag \\
&&-\frac{3\left( 12a^{2}\Delta m-12\overline{q}\Delta q+\widetilde{\kappa }%
^{2}a^{6}\Delta \widetilde{\Lambda }\right) }{4\widetilde{\kappa }%
^{4}a^{9}\!\left( \rho +\lambda \right) ^{2}}\Delta \left( \beta \dot{v}%
^{2}\right)  \notag \\
&&+\frac{3\Delta _{2}}{2\widetilde{\kappa }^{4}a^{12}\left( \rho +\lambda
\right) ^{3}}~.  \label{R_V}
\end{eqnarray}%
with%
\begin{eqnarray}
\Delta _{2} &=&18a^{4}\left( p-\lambda \right) \left( \Delta m\right) ^{2} 
\notag \\
&&+12a^{2}\left( \rho -3p+4\lambda \right) \overline{q}\Delta q\Delta m 
\notag \\
&&+\widetilde{\kappa }^{2}a^{8}\left( 2\rho +3p-\lambda \right) \Delta 
\widetilde{\Lambda }\Delta m  \notag \\
&&-6\left( 2\rho -3p+5\lambda \right) \overline{q}^{2}\left( \Delta q\right)
^{2}  \notag \\
&&-\widetilde{\kappa }^{2}a^{6}\left( \rho +3p-2\lambda \right) \overline{q}%
\Delta q\Delta \widetilde{\Lambda }  \notag \\
&&+\frac{\widetilde{\kappa }^{4}}{24}a^{12}\left( 4\rho +3p+\lambda \right)
\left( \Delta \widetilde{\Lambda }\right) ^{2}
\end{eqnarray}

The Friedmann and Raychaudhuri equations can be derived in an independent
way by adding together all contributions to $U$ and $\Lambda $. To see this,
we derive from the definition of the jump and by employing Eq. (\ref{aveA1})%
\begin{equation}
\Delta \!\left( \frac{A}{B}\right) =\frac{\Delta A}{\overline{B}}-\overline{%
\left( \frac{A}{B}\right) }\frac{\Delta B}{\overline{B}}~.
\end{equation}%
Then, starting from Eq. (\ref{Uembedding}) we compute the detailed
expression of $U^{emb}$, as follows 
\begin{eqnarray}
\kappa ^{2}U^{emb}\! &=&\frac{\!9\left( 12a^{2}\Delta m-12\overline{q}\Delta
q+\widetilde{\kappa }^{2}a^{6}\Delta \widetilde{\Lambda }\right) }{8%
\widetilde{\kappa }^{4}a^{9}\left( \rho +\lambda \right) ^{2}}\Delta \left(
\beta \dot{v}^{2}\right)  \notag \\
&&+\frac{\!9\delta _{2,U}}{8\widetilde{\kappa }^{4}a^{12}\left( \rho
+\lambda \right) ^{3}}~,  \label{Uemb_V}
\end{eqnarray}%
with%
\begin{eqnarray}
\delta _{2,U} &=&12a^{4}\left( \rho -3p+4\lambda \right) \left( \Delta
m\right) ^{2}  \notag \\
&&-24a^{2}\left( 2\rho -3p+5\lambda \right) \overline{q}\Delta q\Delta m 
\notag \\
&&-2\widetilde{\kappa }^{2}a^{8}\left( \rho +3p-2\lambda \right) \Delta
m\Delta \widetilde{\Lambda }  \notag \\
&&+36\left( \rho -p+2\lambda \right) \overline{q}^{2}\left( \Delta q\right)
^{2}  \notag \\
&&+6\widetilde{\kappa }^{2}a^{6}\left( p-\lambda \right) \overline{q}\Delta
q\Delta \widetilde{\Lambda }  \notag \\
&&-\frac{\widetilde{\kappa }^{4}}{4}a^{12}\left( \rho +p\right) \left(
\Delta \widetilde{\Lambda }\right) ^{2}~.
\end{eqnarray}%
Now we can add up the various contributions to $U$, Eqs. (\ref{UWeyl}), (\ref%
{U5D}) and (\ref{Uemb_V}) finding%
\begin{eqnarray}
\kappa ^{2}U &=&\frac{6\overline{m}}{a^{4}}-\frac{9}{2a^{6}}\left( \overline{%
q}^{2}+\frac{\left( \Delta q\right) ^{2}}{4}\right) +\frac{3\overline{\beta 
\dot{v}^{2}}}{2a^{3}}  \notag \\
&&+\frac{\!9\left( 12a^{2}\Delta m-12\overline{q}\Delta q+\widetilde{\kappa }%
^{2}a^{6}\Delta \widetilde{\Lambda }\right) }{8\widetilde{\kappa }%
^{4}a^{9}\left( \rho +\lambda \right) ^{2}}\Delta \left( \beta \dot{v}%
^{2}\right)  \notag \\
&&+\frac{\!9\delta _{2,U}}{8\widetilde{\kappa }^{4}a^{12}\left( \rho
+\lambda \right) ^{3}}~,  \label{U_V}
\end{eqnarray}%
The function $\Lambda $ can be computed, starting from its definition, Eq. (%
\ref{Lambdadef}). We find%
\begin{eqnarray}
\Lambda &=&\Lambda _{0}+\frac{3}{2a^{6}}\left( \overline{q}^{2}+\frac{\left(
\Delta q\right) ^{2}}{4}\right) -\frac{3\overline{\beta \dot{v}^{2}}}{2a^{3}}
\notag \\
&&-\frac{9\left( 12a^{2}\Delta m-12\overline{q}\Delta q+\widetilde{\kappa }%
^{2}a^{6}\Delta \widetilde{\Lambda }\right) }{8\widetilde{\kappa }%
^{4}a^{9}\left( \rho +\lambda \right) ^{2}}\Delta \left( \beta \dot{v}%
^{2}\right)  \notag \\
&&+\frac{9\delta _{2,\Lambda }}{8\widetilde{\kappa }^{4}a^{12}\left( \rho
+\lambda \right) ^{3}}\ ,  \label{La_V}
\end{eqnarray}%
with%
\begin{eqnarray}
\delta _{2,\Lambda } &=&12a^{4}\left( \rho +3p-2\lambda \right) \left(
\Delta m\right) ^{2}  \notag \\
&&-72a^{2}\left( p-\lambda \right) \overline{q}\Delta q\Delta m  \notag \\
&&+6\widetilde{\kappa }^{2}a^{8}\left( \rho +p\right) \Delta m\Delta 
\widetilde{\Lambda }  \notag \\
&&-12\left( \rho -3p+4\lambda \right) \overline{q}^{2}\left( \Delta q\right)
^{2}  \notag \\
&&-2\widetilde{\kappa }^{2}a^{6}\left( 2\rho +3p-\lambda \right) \overline{q}%
\Delta q\Delta \widetilde{\Lambda }  \notag \\
&&+\frac{\widetilde{\kappa }^{4}}{12}a^{12}\left( 5\rho +3p+2\lambda \right)
\left( \Delta \widetilde{\Lambda }\right) ^{2}~.
\end{eqnarray}%
Inserting $U$ and $\Lambda $ in Eqs. (\ref{Fried}) and (\ref{Raych}) we
recover the explicit form of the Friedmann and Raychaudhuri equations
derived earlier in this section, Eqs. (\ref{Fr_V}) and (\ref{R_V}). In the
process we use%
\begin{align}
\delta _{2,\Lambda }+\delta _{2,U}& =\frac{2\widetilde{\kappa }^{4}}{3}%
a^{10}\left( \rho +\lambda \right) ^{3}\left( \Delta B\right) ^{2}~,
\label{dep} \\
\delta _{2,\Lambda }-\delta _{2,U}& =4\Delta _{2}~.  \label{dem}
\end{align}%
Thus we have the complete set of dynamical equations for the Friedmann brane
with variable tension, embedded in charged VAdS5 space-time.

\subsection{The consistency of the model}

In Section III we argued that the twice-contracted Bianchi equation emerges
as consequence of the Friedmann, Raychaudhuri and energy balance equations,
provided they hold independently. This was a consequence of the effective
energy density $U$ depending in an unspecified way on the embedding, 5D
sources and non-local sources of the gravitational field (Weyl fluid). In
this Section however we have explicitly constructed the embedding, and
assumed that the 5D matter consists of a charged radiation field propagating
in interior or exterior pieces of charged VAdS5 space-time. As such, the
system of Friedmann, Raychaudhuri and energy balance equations become
inter-related, as in general relativity.

In order to illustrate this, we restrict ourselves to a particularly simple
case, with symmetric embedding, no electric charge, and flat spatial
sections $k=0$. We will prove for this case, that the Raychaudhuri equation
emerges from the Friedmann and the energy balance equations in the same way
as in general relativity. In the process we will be able to remove the sign
ambiguity represented by $S_{1}$ and $S_{2}$ and make further comments on
the validity of the model discussed in this Section.

The energy balance equation (\ref{conti_V}) simplifies to%
\begin{equation}
\dot{\rho}+\dot{\lambda}+3\frac{\dot{a}}{a}\left( \rho +p\right) =\epsilon
\left( -1\right) ^{\eta }\frac{6\beta \dot{v}^{2}}{\widetilde{\kappa }%
^{2}a^{3}}\ .  \label{rhodot}
\end{equation}%
As noted before, the global sign $\epsilon \left( -1\right) ^{\eta }$ is
negative when the brane radiates away energy (for example by thermic
radiation) and positive, if radiation is absorbed by the brane (such a
radiation can be emitted by a 5D black hole, if the 5D region is an interior
patch of VAdS5, or may come from the 5D infinity, for exterior regions).

The brane bounds a region of VAdS5 characterized by the mass function $%
m\left( v\right) $ (or equivalently $m\left( t\right) $, when seen from the
brane). When the brane radiates away energy into the 5D space-time, this
mass function should increase in $\tau $, irrespective of the movement of
the brane. As this movement is sub-luminal, the brane will never pass its
own emitted radiation, even if it moves in the same direction. When the brane absorbs radiation, such that the total energy of
the bounded 5D space-time region decreases, the mass $m\left( \tau \right) $
will become smaller. In consequence $sgn\left( \dot{m}\right) =-\epsilon
\left( -1\right) ^{\eta }$.

Let us now derive an expression for $\dot{m}$. We start from Eq. (\ref{beta}%
), which can be rewritten as $\dot{m}=\epsilon \beta \dot{v}$. By
remembering that $sgn\left( \dot{v}\right) =\,S_{1}$ and that $\beta >0$ for
radiation obeying the energy conditions\footnote{%
For a discussion of exotic radiation, corresponding to $\beta <0$ in 4D, see
Ref. \cite{wormhole}.}, we have that $sgn\left( \dot{m}\right) =\epsilon
S_{1}$. Thus we conclude that $S_{1}=\left( -1\right) ^{\eta +1}$.

Under the simplifying assumptions of this subsection the Friedmann equation
is%
\begin{equation}
\frac{\dot{a}^{2}}{a^{2}}=\frac{\kappa ^{2}}{6\lambda }\!\left( \rho
+\lambda \right) ^{2}+\frac{2m}{a^{4}}+\frac{\widetilde{\kappa }^{2}%
\widetilde{\Lambda }}{6}~.  \label{Friedmann_simple}
\end{equation}%
It's time derivative, employing the energy-balance equation (\ref{rhodot})
gives:%
\begin{eqnarray}
\frac{\dot{a}}{a}\frac{\ddot{a}}{a} &=&\frac{\dot{a}}{a}\left\{ \frac{%
\Lambda _{0}}{3}\!-\frac{\kappa ^{2}}{6}\!\left[ \rho \left( 1+\frac{2\rho }{%
\lambda }\right) +3p\left( 1+\frac{\rho }{\lambda }\right) \right] -\frac{2m%
}{a^{4}}\right\}  \notag \\
&&+\frac{\dot{m}}{a^{4}}+\epsilon \left( -1\right) ^{\eta }\frac{\widetilde{%
\kappa }^{2}}{6}\!\left( \rho +\lambda \right) \frac{\beta \dot{v}^{2}}{a^{3}%
}~.  \label{next_to_last}
\end{eqnarray}%
By employing Eqs. (\ref{vdotm1}) and (\ref{Friedmann_simple}) we obtain%
\begin{equation}
\dot{m}=\frac{\epsilon \beta \dot{v}^{2}}{\dot{v}}=\beta \dot{v}^{2}a\left[ -%
\frac{\dot{a}}{a}+\epsilon \left( -1\right) ^{\eta +1}\frac{\widetilde{%
\kappa }^{2}}{6}\!\left( \rho +\lambda \right) \right] ~.  \label{mdot}
\end{equation}%
The last two terms of Eq. (\ref{next_to_last}), after inserting Eq. (\ref%
{mdot}), reduce to $-\left( \beta \dot{v}^{2}/a^{3}\right) \left( \dot{a}%
/a\right) $. After simplifying with $\dot{a}/a$ we recover the Raychaudhuri
equation:%
\begin{equation}
\frac{\ddot{a}}{a}=\frac{\Lambda _{0}}{3}-\!\frac{\kappa ^{2}}{6}\!\left[
\rho \left( 1+\frac{2\rho }{\lambda }\right) +3p\left( 1+\frac{\rho }{%
\lambda }\right) \right] -\frac{2m}{a^{4}}-\frac{\beta \dot{v}^{2}}{a^{3}}~.
\end{equation}

Having no other constraint on the rest of the signs, we can choose $%
S_{2}=\epsilon $, so that Eq. (\ref{vprime}) agrees with the corresponding
equation in Ref. \cite{Decomp}.

\section{Discussion and Concluding Remarks}

In this paper we have introduced the possibility of a variable brane tension
in the context of co-dimension one brane-worlds in which both the brane and
the 5D space-time are curved. This possibility has not been explored before, although there is no a priori reason, why the brane tension should be independent of temperature, while the tension of the fluid membranes definitely is. The evolution
of the brane tension triggers a variable gravitational constant (through Eq.
(\ref{kappa2def})) and a variable cosmological constant (through Eq. (\ref%
{Lambda0})).

The sources for the curvature in the model are (a) non-standard model fields
and a negative cosmological constant in 5D, and (b) distributional standard
model sources on the brane, together with the brane tension. We have considered this
setup without any particular symmetry assumption or further specification of
the sources. Gravitational dynamics regarded from a brane point of view is
expressed by an effective Einstein equation, the Codazzi equation and the
twice-contracted Gauss equation. Beyond the distributional standard model
sources already mentioned and the effective 4D cosmological constant, other
sources appear in the effective Einstein equation, which are a quadratic
source term (arising by replacing the quadratic expressions in the extrinsic
curvature with matter terms by use of the junction conditions); a Weyl fluid
term (originating in the 5D Weyl curvature); an asymmetry source term (from
the possible asymmetry of the embedding); and an 5D matter term (from the
pull-back of the 5D sources). Various other equations were derived,
expressing either constraints on the embedding and 5D matter sources or the
off-brane evolution of the Weyl fluid. The Lanczos equation and Bianchi
identity were also given. A careful analysis has identified those equations,
which exhibit a $\dot{\lambda}$ term: these are the twice-contracted Bianchi
identity and the Codazzi relation. The manifest form of the other equations
is not changed, although their solutions will depend on the specific way the
brane tension varies.

Then we have specified the formalism for a cosmological context, discussing
Friedmann branes containing perfect fluid. The Codazzi equation generates an
energy-balance equation, expressing the possible energy interchange between
the brane and the 5D sources and also depends on the time derivative of the
brane tension. (In the absence of such an interchange and for constant brane
tension the fluid obeys the usual continuity equation.)\ When the collection
of Weyl, asymmetry and 5D matter sources obey the cosmological symmetries,
they are characterized by a time-dependent potential. The time evolution of
this potential is given by the twice-contracted Bianchi identity.

Next we have considered the most generic 5D space-time with cosmological
constant, electromagnetic field and radiation (the VAdS5 space-time), which
has the property that a Friedmann brane can be embedded into it at any
point. The detailed discussion of the 5D geometry led to the expression of
the Weyl fluid source, not given before. The above-mentioned sources obeying
the 5D Einstein equation gave the pull-back 5D matter source term and the
respective contribution to the 4D cosmological constant. The analysis of the
embedding led to the asymmetry source term and the asymmetry contribution to
the 4D cosmological constant. These are new results. The explicit knowledge
of various contributions to the potential $U$ will allow to easily apply the
formalism for special cases, when either the mass, the cosmological constant
or the charge is symmetric, or the masses / charges are missing from the 5D
space-time. Gravitational dynamics on the brane was derived in the form of
the generalized Friedmann, generalized Raychaudhuri and energy-balance
equations, given explicitly. From among these, only the energy balance
equation has a manifest $\dot{\lambda}$-dependence. We have verified these
results by an independent derivation of the generalized Friedmann and
Raychaudhuri equations.

Finally we have shown in the special case without charge, symmetric
embedding and spatially flat sections how the Raychaudhuri equation emerges
from the derivative of the Friedmann equation, the energy-balance equation
and the expression of $\dot{m}$. This derivation closely resembles the
general relativistic result. In the process we have fixed an existing sign
ambiguity: $S_{1}=\left( -1\right) ^{\eta +1}$. In the derivation of the
dynamics we have assumed the same $S_{1}$ on both sides of the brane, therefore our
results hold for a brane, which is the common boundary of either two
interior ($S_{1}=1$) or two exterior ($S_{1}=-1$) regions, whereas the
results of \cite{Decomp} hold only for the matching of two interior regions,
as in that paper the positive sign was chosen in Eq. (\ref{vdot}).\ (All
applications to specific models of the formalism derived in \cite{Decomp}
were for a brane bounding interior regions \cite{RadBrane}, \cite{GK}, \cite%
{KKG}.) The cases characterized by $S_{1}^{L}S_{1}^{R}=-1$ are not
incorporated in the present analysis, their discussion in full detail being
beyond the scope of the present paper.

With these the formalism is set for analyzing particular brane-world models
with variable tension. In order to do this in a cosmological context, we
need the scale-factor dependence of the brane tension. As one of the generic
features of the models with temperature-dependent tension is that the 4D
gravitational 'constant' (defined in terms of the brane tension) evolves in
time, there may be certain similarities with other models with variable
gravitational 'constant' (for some recent proposals see \cite{varG}, \cite%
{varG2}).

In a parallel work \cite{VarBraneTensionPRL} we have investigated the flat E%
\"{o}tv\"{o}s brane embedded symmetrically in an uncharged VAdS5 space-time.
Such a brane exhibits the E\"{o}tv\"{o}s law for the temperature dependence
of the brane. By additionally imposing that the cosmological perfect fluid
obeys the continuity equation, the dynamics was considerably simplified. The
continuity equation can be obeyed by fine-tuning the energy interchange
between the brane and 5D radiation with the evolution of the brane tension.
In a reasonable parameter range the emerging cosmology is open, reproducing
a decelerated expansion followed by an accelerated phase. In the process the
mass of the VAdS5 regions decreases until the 5D radiation is completely
fueled out, such that the 5D space-time becomes Anti de Sitter. Both the
brane tension and the 4D cosmological constant evolve from infinitesimal
values at the formation of the brane in a very hot early universe towards
late-time constant values. The 4D cosmological constant evolves from huge
negative values, contributing to gravitational attraction in the early
universe, to a small positive value. Cosmological expansion then continues
in a de Sitter phase.

The next simplest model with variable tension would arise for a static 5D
space-time. Without energy interchange between the brane and 5D space-time,
the energy-balance equation becomes a continuity equation for the perfect
fluid with energy density $\rho +\lambda $ and pressure $p-\lambda $. A
detailed investigation of this model is in progress.

Various other models with temperature-dependent brane tension can be
constructed in the framework of the formalism developed in the present
paper, by considering specific radiation fields in VAdS5, or adopting a
temperature dependence, which is different from the E\"{o}tv\"{o}s' law
established for fluid membranes.

\section{Acknowledgments}

This work was supported by the OTKA grant 69036, the J\'{a}nos Bolyai Grant
of the Hungarian Academy of Sciences and a London South Bank University
Research Opportunities Fund.


\begin{thebibliography}{99}
\bibitem{RS2} L. Randall, R. Sundrum, \textit{Phys. Rev. Lett.} \textbf{83,}
4690 (1999).

\bibitem{SMS} T. Shiromizu, K.I. Maeda, M. Sasaki, \textit{Phys. Rev.} D 
\textbf{62,} 024012 (2000).

\bibitem{MaartensLivRev} R. Maartens, \textit{Living Rev. Rel.} \textbf{7}, 7 (2004).

\bibitem{Decomp} L.\'{A}. Gergely, \textit{Phys. Rev.} D \textbf{68}, 124011 
\textit{(}2003). [Errata: In the last term of Eq. (37) the summation indices 
$ab$ should be replaced by $cd$. On the left hand side of Eq. (67) the
indices $cd$ should be replaced by $ab$. 
In Eq. (62) $\dot v$ should be replaced by $\dot t$.
In Eqs. (82), (83), (85) and (86) $%
\overline{q^{2}}$ should be replaced by $\overline{q}^{2}$. The last terms
of Eqs. (68) and (99) should have $\widetilde{\kappa }q$ in place of $q$%
. In Eq. (102) the factor $2$ from the denominator of the last term should
be removed. The sentence preceding Eq. (103) should read "The Raychaudhuri
equation acquires two new terms on the right hand side." ] The conversion to
the notations of the present paper is $( l,~\widetilde{T}_{ab},~%
\widetilde{\Pi }_{ab},~\widetilde{\kappa }q) \rightarrow ( y,~-%
\widetilde{\Lambda }\widetilde{g}_{ab}+\widetilde{T}_{ab}+\tau _{ab}\delta
\left( y\right) ,~-\widetilde{\Lambda }\widetilde{g}_{ab}+\widetilde{T}%
_{ab},~q) $. In particular in this paper $\widetilde{T}_{ab}$ denotes
the energy-momentum tensor of the 5D fields, while there it denoted the
total 5D energy-momentum tensor, which included the cosmological constant
and the distributional contribution.

\bibitem{asymmetry} P. Kraus, \textit{J. High Energy Phys. }\textbf{9912},
011 (1999).

D. Ida, \textit{J. High Energy Phys. }\textbf{0009}, 014 (2000).

A.C. Davis, I. Vernon, S.C. Davis, W.B. Perkins, \textit{Phys. Letters }B 
\textbf{504}, 254 (2001).

N. Deruelle, T. Dole\v{z}el, \textit{Phys. Rev.} D \textbf{62}, 103502 (2000)

W.B. Perkins, \textit{Physics Lett. }B \textbf{504}, 28 (2001).

B. Carter, J.-P. Uzan, \textit{Nucl. Phys. }B \textbf{606}, 45 (2001).

H. Stoica, H. Tye, I. Wasserman, \textit{Phys. Lett. }B \textbf{482}, 205
(2000).

\bibitem{Israel} W. Israel, \textit{Nuovo Cimento B} \textbf{44}, 1 (1966); erratum: B \textbf{49}, 463 (1967).

\bibitem{Lanczos} C. Lanczos, \textit{Phys. Zeils.}, \textbf{23}, 539
(1922); \textit{Ann. der Phys.}\textbf{\ 74}, 518 (1924).

\bibitem{tidalRN} N. Dadhich, R. Maartens, P. Papadopoulos, V. Rezania, 
\textit{Phys. Lett} B \textbf{487}, 1 (2000).

\bibitem{Aliev} A.N. Aliev, \ A.E. Gumrukcuoglu, \textit{Phys. Rev. D} 
\textbf{71}, 104027 (2005).

\bibitem{BraneSwissCheese} L.\'{A}. Gergely, \textit{Phys. Rev.} D \textbf{74%
}, 024002 (2006).

L.\'{A}. Gergely, I. K\'{e}p\'{\i}r\'{o}, \textit{JCAP} \textbf{07} (07),
007 (2007).

\bibitem{BGM} M. Bruni, C. Germani, R. Maartens, \textit{Phys. Rev. Lett. }%
\textbf{87}, 231302 (2001).

\bibitem{DadhichGhosh} N. Dadhich, S.G. Ghosh, \textit{Physics Letters }B 
\textbf{518}, 1 (2001).

\bibitem{GovenderDadhich} M. Govender, N. Dadhich, \textit{Physics Letters }%
B \textbf{538}, 233 (2002).

\bibitem{CasadioGermani} R. Casadio, C. Germani, \textit{Prog. Theor. Phys.} 
\textbf{114}, 23 (2005).

\bibitem{Pal} S. Pal, \textit{Phys. Rev. }D \textbf{74}, 124019 (2006).

\bibitem{BraneOppSnyder} L.\'{A}. Gergely, \textit{JCAP }\textbf{07}02, 027
(2007).

\bibitem{GM} C. Germani, R. Maartens, \textit{Phys. Rev. }D \textbf{64},
124010 (2001).

\bibitem{ND} N. Deruelle, \textit{Stars on branes: the view from the brane,\ 
}gr-qc/0111065 (2001).

\bibitem{Ovalle} J. Ovalle, \textit{Searching exact solutions for compact
stars in braneworld: A Conjecture, }gr-qc/0703034 (2007), to be published in 
\textit{Mod. Phys. Letters A.}

\bibitem{HarkoRC} M.K. Mak, T. Harko, \textit{Phys.Rev. D }\textbf{70},
024010 (2004).

T. Harko, K.S. Cheng, \textit{Astrophys.J.} \textbf{636}, 8-20 (2006).

C.G. Boehmer, T. Harko, \textit{Class. Quantum Grav.} \textbf{24}, 3191
(2007).

\bibitem{HarkoClusters} T. Harko, K.S. Cheng, \textit{Phys. Rev. D} \textbf{%
76}, 044013 (2007).

\bibitem{GeDa} L.\'{A}. Gergely, B. Dar\'{a}zs, \textit{Publ. Astron. Dept. E%
\"{o}tv\"{o}s Univ. PADEU} \textbf{17}, 213 (2006); astro-ph/0602427.

\bibitem{BohmerHarkoLobo} C.G. Boehmer, T. Harko, F.S.N. Lobo,\ \textit{%
Class. Quant. Grav.} \textbf{25}, 045015 (2008).

\bibitem{BDEL} P Bin\'{e}truy, C Deffayet, U Ellwanger, D Langlois, \textit{%
Phys. Lett.} B \textbf{477}, 285 (2000).

\bibitem{ChKN} A. Chamblin, A. Karch, A. Nayeri, \textit{Phys. Lett. }B, 
\textbf{509}, 163 (2001).

\bibitem{PalStructure} S. Pal, \textit{Phys. Rev. D }\textbf{74}, 024005
(2006).

S. Pal, \textit{Phys. Rev. D }\textbf{78}, 043517
(2008).

\bibitem{perturb} C. de Rham, \textit{Phys. Rev. D} \textbf{71}, 024015
(2005).

K. Koyama, A. Mennim, V.A. Rubakov, D. Wands, T.Hiramatsu, \textit{JCAP} 
\textbf{07}04, 001 (2007).

C. de Rham, S. Watson, \textit{Class. Quant. Grav.} \textbf{24}, 4219 (2007).

\bibitem{Nucleosynthesis} J.D. Bratt, A.C. Gault, R.J. Scherrer, T.P.
Walker, \textit{Phys. Lett B} \textbf{546}:19 (2002).

\bibitem{supernova} Z. Keresztes, L.\'{A}. Gergely, B. Nagy, G.M. Szab\'{o}, 
\textit{PMC Physics A} \textbf{1} : 4 (2007).

G.M. Szab\'{o}, L.\'{A}. Gergely, Z. Keresztes, \textit{PMC Physics A} 
\textbf{1} : 8 (2007).

L.\'{A}. Gergely, Z. Keresztes, G.M. Szab\'{o}, \textit{AIP Conference
Proceedings} \textbf{957}, 391 (2007).

\bibitem{tabletop} J.C. Long et al., \textit{Nature} \textbf{421}, 922
(2003).

\bibitem{nucleosynthesis} R. Maartens, D. Wands, B.A. Bassett, I.P.C. Heard, 
\textit{Phys. Rev. D} \textbf{62}, 041301(R) (2000)

\bibitem{GK} L.\'{A}. Gergely, Z. Keresztes, \textit{JCAP} \textbf{06 }(01),
022 (2006).

\bibitem{Eotvos} R. E\"{o}tv\"{o}s, \textit{Wied. Ann.} \textbf{27}, 448
(1886).

\bibitem{BCG} P Bowcock, C Charmousis, R Gregory, \textit{Class. Quantum
Grav.} \textbf{17}, 4745 (2000).

\bibitem{BraneBlackHole} R. Gregory, \textit{Braneworld Black Holes, }arXiv:
0804.2595 [hep-th] (2008).

\bibitem{EinBrane} L.\'{A}. Gergely, R. Maartens, \textit{Class. Quantum
Grav.} \textbf{19}, 213 (2002).

\bibitem{EinBrane3} Z. Keresztes, L.\'{A}. Gergely, \textit{Class. Quantum
Grav.} \textbf{25}, 165016 (2008).

\bibitem{LSR} D. Langlois, L. Sorbo, M Rodr\'{\i}guez-Mart\'{\i}nez, \textit{%
Phys. Rev. Lett.}{\ }\textbf{89}, 171301 (2002).

\bibitem{VernonJennings} D. Jennings, I.R. Vernon, \textit{JCAP} \textbf{05}%
07, 011 (2005).

\bibitem{Langlois} D. Langlois, \textit{Prog.Theor.Phys.Suppl.} \textbf{163}%
, 258 (2006).

\bibitem{Jennings} D. Jennings, I.R. Vernon, A.-C. Davis, C. van de Bruck, 
\textit{JCAP} \textbf{05}04, 013 (2005).

\bibitem{RadBrane} L.\'{A}. Gergely, E. Leeper, and R. Maartens, \textit{%
Phys.Rev.} D \textbf{70}, 104025 (2004).

\bibitem{KKG} Z. Keresztes, I. K\'{e}p\'{\i}r\'{o}, and L.\'{A}. Gergely, 
\textit{JCAP} \textbf{06}05, 020 (2006).

\bibitem{HarkoInflation} T. Harko, W.F. Choi, K. C. Wong, K.S. Cheng, 
\textit{JCAP} \textbf{08}06 002 (2008).

\bibitem{wormhole} L.\'{A}. Gergely, \textit{Phys. Rev.} D \textbf{65},
127503 \textit{(}2002).

\bibitem{varG} J.D. Barrow, J. Magueijo, H.B. Sandvik, \textit{Phys.Lett. B} 
\textbf{541} 201 (2002).

\bibitem{varG2} J.D. Bekenstein, E. Sagi, \textit{Do Newton's G and
Milgrom's a\_0 vary with cosmological epoch?, } arXiv:0802.1526v1 [astro-ph]
(2008).

\bibitem{VarBraneTensionPRL} L.\'{A}. Gergely, \textit{E\"{o}tv\"{o}s branes}%
, arxiv:0806.4006 [gr-qc] (2008).
\end{thebibliography}
\end{document}